\documentclass[aps,prb,twocolumn,superscriptaddress]{revtex4} 
\usepackage{graphicx}
\usepackage{color}
\usepackage{todonotes}
\begin{document}

\title{Magnetic excitations in stripe-ordered La$_{1.875}$Ba$_{0.125}$CuO$_4$ studied using resonant inelastic x-ray scattering}

\author{M. P. M. Dean}
\email{mdean@bnl.gov}
\affiliation{Department of Condensed Matter Physics and Materials Science,Brookhaven National Laboratory, Upton, New York 11973, USA}

\author{G. Dellea}
\author{M. Minola}
\affiliation{CNR/SPIN, CNISM and Dipartimento di Fisica, Politecnico di Milano, piazza Leonardo da Vinci 32, 20133 Milano, Italy}

\author{S. B. Wilkins}
\author{R. M. Konik}
\author{G. D. Gu}
\affiliation{Department of Condensed Matter Physics and Materials Science,Brookhaven National Laboratory, Upton, New York 11973, USA}

\author{M. Le Tacon}
\affiliation{Max-Planck-Institute for Solid State Research, D-70569 Stuttgart, Germany}

\author{N. B. Brookes}
\author{F. Yakhou-Harris}
\author{K. Kummer}
\affiliation{European Synchrotron Radiation Facility (ESRF), BP 220, F-38043 Grenoble Cedex, France}

\author{J. P. Hill}
\affiliation{Department of Condensed Matter Physics and Materials Science,Brookhaven National Laboratory, Upton, New York 11973, USA}

\author{L. Braicovich}
\author{G. Ghiringhelli}
\email{giacomo.ghiringhelli@polimi.it}
\affiliation{CNR/SPIN, CNISM and Dipartimento di Fisica, Politecnico di Milano, piazza Leonardo da Vinci 32, 20133 Milano, Italy}

\def\mathbi#1{\ensuremath{\textbf{\em #1}}}
\def\Q{\ensuremath{\mathbi{Q}}}
\newcommand{\angstrom}{\mbox{\normalfont\AA}}

\date{\today}

\begin{abstract}
The charge and spin correlations in La$_{1.875}$Ba$_{0.125}$CuO$_4$ (LBCO 1/8) are studied using Cu $L_3$ edge resonant inelastic x-ray scattering (RIXS). The static charge order (CO) is observed at a wavevector of $(0.24,0)$ and its charge nature confirmed by measuring the dependence of this peak on the incident x-ray polarization. The paramagnon excitation in LBCO 1/8 is then measured as it disperses through the CO wavevector. Within the experimental uncertainty no changes are observed in the paramagnon due to the static CO, and the paramagnon seems to be similar to that measured in other cuprates, which have no static CO. Given that the stripe correlation modulates both the charge and spin degrees of freedom, it is likely that subtle changes do occur
in the paramagnon due to CO. Consequently, we propose that future RIXS measurements, realized with higher energy resolution and sensitivity, should be performed to test for these effects.
\end{abstract}


\maketitle

The copper-oxide superconductors play host to strong correlations between their charge, lattice and spin degrees of freedom, which in many cuprates drives the formation of modulations or ``stripes'' of charge order (CO), spin order (SO) and lattice order.\cite{Kivelson2003, Vojta2009, Fujita2012stripes} In certain cuprates including La$_{2-x}$Ba$_x$CuO$_4$ and La$_{1.6-x}$Sr$_x$Nd$_{0.4}$CuO$_4$ these stripes are stabilized into static modulations of the charge and spin order, where the relationship between CO and SO is reinforced by the fact that the CO incommensurability is half that of the SO incommensurability over a range of different dopings.
\cite{Tranquada1995, Tranquada1997, Ichikawa2000, Fujita2004, Hucker2010}
The static SO also affects the dynamic magnetic properties, and inelastic neutron scattering observes that magnetic excitations emanate from the magnetic SO Bragg peaks. \cite{Tranquada2004, Fujita2004, Xu2007} In the nickelate La$_{5/3}$Sr$_{1/3}$NiO$_4$ strong stripe correlations are also found, and here, new magnetic excitations appear upon cooling into the stripe-ordered phase, which were interpreted as dynamic quasi 1D stripe correlations,\cite{Boothroyd2003} implying that similar excitations might exist in the cuprates.

There is considerable evidence that stripes in the cuprates are intimately related to high-$T_c$ superconductivity. In La$_{2-x}$Ba$_x$CuO$_4$ the doping levels for which static stripe order is stabilized, at $x\approx 1/8$, correspond to a suppression of superconducting $T_c$ -- a phenomenon known as the $1/8$ anomaly. \cite{Moodenbaugh1988} While static stipe order appears to suppress superconductivity, some researchers have suggested that dynamic stripe fluctuations may act to promote superconductivity.\cite{Emery1997, Kivelson1998} There is also debate as to whether the spin, charge or lattice degrees of freedom are most important factor for causing superconducting pairing. Spin fluctuation mediated pairing is perhaps the most intensely studied scenario for high-$T_c$ superconductivity, \cite{Eschrig2006, Scalapino2012} but these theories compete with ideas based on charge fluctuations.\cite{Castellani1995} All these proposals imply that a good understanding of the charge and spin correlations in the cuprates is a prerequisite to solving the long-standing problem of high-$T_c$ superconductivity. In particular, we must understand how static charge and spin order in the cuprates can affect the dynamic charge and spin correlations.

 Driven primarily by improvements in experimental resolution,\cite{Ghiringhelli2004} Cu $L_3$ edge RIXS now provides a powerful probe for these correlations -- for example, RIXS had an important role in the discovery of CO in YBa$_2$Cu$_3$O$_{6+x}$.\cite{Ghiringhelli2012} RIXS is also sensitive to magnetic excitations around the CO wavevector $(0.24, 0)$, where the magnetic excitations are very weak relative to those around the SO wavevector $(0.38, 0.5)$ and consequently are difficult to measure using current inelastic neutron scattering techniques.

Here we report RIXS measurements of the magnetic excitations in stripe-ordered LBCO 1/8 as they disperse through the CO wavevector, complementing inelastic neutron scattering studies of the magnetic excitations near the SO wavevector.\cite{Tranquada2004, Kivelson2003, Vojta2009}
We start by measuring the CO stripe at $(0.24,0)$ which is shown to follow the expected incident x-ray polarization dependence for charge (rather than spin) scattering. This CO peak  is tracked up to 65~K, before it disappears at 85~K. The magnetic excitations in LBCO 1/8 are then examined as they disperse through the CO wavevector. Within the accuracy of the current measurements, the high-energy paramagnon is almost unaffected by the CO. Rather, the paramagnon appears to be analogous to that observed at low temperatures in other cuprates with no static CO.\cite{Braicovich2009, Braicovich2010, BraicovichPRB2010, LeTacon2011, Dean2012, Dean2013a, LeTacon2013, Dean2013b} We suggest that future higher energy resolution RIXS experiments should measure the lower energy magnetic excitations in LBCO which are more likely to be affected by interactions with the CO.

An La$_{1.875}$Ba$_{0.125}$CuO$_4$ (LBCO 1/8) single crystal was grown at Brookhaven National Laboratory using the floating zone method and shown to be of high quality in previous soft x-ray studies.\cite{Wilkins2011, Thampy2013} Throughout this manuscript wavevectors will be described using the high temperature tetragonal ($I4/mmm$) space group with $a=b=3.78$~\angstrom{} and $c=13.28$~\angstrom{}. The sample was cleaved \emph{ex situ} to reveal a face with a $[001]$ surface normal and mounted with the $[100]$ and $[001]$ directions in the scattering plane.

Cu $L_3$ edge RIXS measurements were performed using the AXES instrument at the ID08 beamline at the European Synchrotron Radiation Facility.\cite{Ghiringhelli2006_AXES,Dallera1996} 
The incident x-ray energy was set to the peak in the measured Cu $L_3$ edge x-ray absorption spectrum and the x-ray polarization was set either parallel ($\pi$) or perpendicular ($\sigma$) to the scattering plane. Experiments were performed with a fixed scattering angle $2\theta=130^{\circ}$ and the sample was rotated about a vertical axis in order to vary $Q_{\parallel}$ and $Q_{\perp}$, the projections of the scattering vector $\Q{}$ along $[100]$ and  $[001]$ respectively. We note that previous studies have shown that the stripe peak is very broad along $L$,\cite{Hucker2013} so the CO peak can be observed even though we are not at the peak in $Q_{\perp}$. Positive \Q{} is defined so that high \Q{} corresponds to x-rays being emitted close to parallel to the sample surface. \cite{supplementary} The combined resolution function of the monochromator and spectrometer is approximately Gaussian with a half- width-half-maximum (HWHM) of 130~meV as determined by measuring the non-resonant elastic scattering from disordered carbon tape.

\begin{figure}
\includegraphics{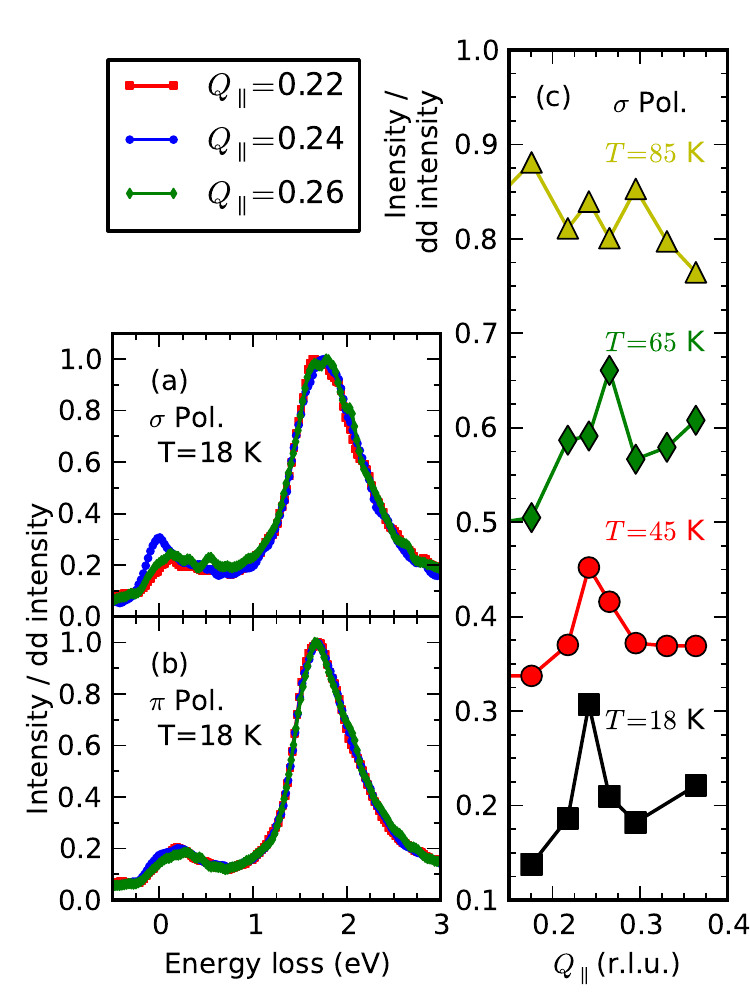} %
\caption{(Color online) (a)\&(b) RIXS spectra of LBCO 1/8 at $T=18$~K for 3 different $Q_{\parallel}$ near the stripe vector at $Q_{\parallel}=0.24$. (a) uses $\sigma$-polarized incident x-rays which enhances charge scattering and (b) uses $\pi$-polarized incident x-ray which enhances magnetic scattering. (c) Temperature dependence of the elastic RIXS intensity as measured with $\sigma$-polarized incident x-rays. Scans are offset vertically for clarity.}
\label{Fig1}
\end{figure}

Figure~\ref{Fig1}(a)\&(b) plots Cu $L_3$ edge RIXS spectra of LBCO 1/8. Between 1 to 3~eV energy loss, the strong $dd$ excitations are visible, corresponding to transitions of the hole in the valence band into higher energy Cu $d$-orbitals.\cite{Ghiringhelli2004} As in previous studies, this excitation is used as a calibration standard in order to compare the intensities of different spectra.\cite{BraicovichPRB2010} We then scan $Q_{\parallel}$, focusing on the low-energy loss region of the spectrum (0-500~meV), which contains information about the static and low-energy spin and charge correlations.\cite{Ament2011} An increase in the low-energy scattering is observed in Fig.~\ref{Fig1}(a) with $\sigma$ polarized incident x-rays at $Q_{\parallel}=0.24$, where the static CO peak is known to exist from other studies \cite{Fujita2004, Abbamonte2005, Kim2008, Tranquada2008, Hucker2011, Wilkins2011, Thampy2013} including those measuring the same sample.\cite{Wilkins2011, Thampy2013} Upon changing the incident x-ray polarization from $\sigma$ in Fig.~\ref{Fig1}(a) to $\pi$ in Fig.~\ref{Fig1}(b) the CO peak is strongly suppressed. In Cu $L_3$ edge studies of the cuprates using this geometry, $\sigma$ polarized incident x-rays enhance the contribution of  charge scattering to the spectrum; while $\pi$ polarized incident x-rays enhance the contribution of magnetic scattering to the spectrum. Such a trend is observed in undoped \cite{Braicovich2009, Braicovich2010, BraicovichPRB2010, Guarise2010, LeTacon2011, Dean2012} and doped cuprates \cite{LeTacon2011, Dean2013a, Dean2013b} as well as being predicted theoretically.\cite{Ament2009, Haverkort2010, Igarashi2011} Thus this polarization dependence confirms the predominantly charge nature of this peak, although these measurements do not distinguish whether the scattering comes from the doped holes or a structural modulation of the position of the Cu atoms.\cite{Abbamonte2005}

\begin{figure}
\includegraphics{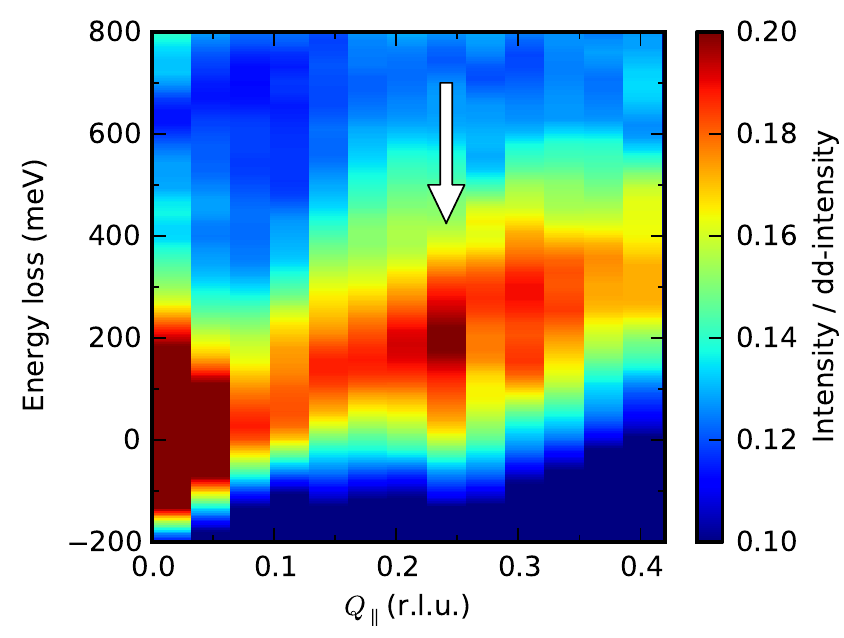} %
\caption{(Color online) RIXS intensity dispersion of LBCO 1/8 showing the dispersion of the paramagnon from low energy at $\Q_{\parallel}=0$ up to $\sim 300$~meV near the Brillouin zone boundary. The white arrow marks the CO wavevector at $\Q_{\parallel}=0.24$. The data were taken at $T=18$~K with $\pi$ polarized incident x-rays and the spectra have been normalized so that the peak intensity of the $dd$-excitations is one.}
\label{Fig2}
\end{figure}

We now examine the temperature evolution of the RIXS intensity at zero energy transfer by performing \Q{} scans with $\sigma$ polarized incident x-rays. It should be noted that this scan tracks the existence of the CO peak, but current soft x-ray RIXS spectrometers do not have the angular freedom required to precisely align the measurement to the peak of the scattering in \Q{}. The intensity of the CO peak plotted in Fig.~\ref{Fig1}(c) is seen to drop with increasing temperature before disappearing in the $T=85$~K scan. Notably, the CO is still visible at 65~K, whereas O K-edge energy-integrated resonant soft x-ray scattering (RSXS) measurements on the same crystal do not observe a peak above 56~K.\cite{Wilkins2011} Our results are, however, compatible with the RSXS results, since the latter observe that the CO peak is broadened with increasing temperature while maintaining roughly constant integrated intensity. 56~K represents the point at which the CO peak becomes too broad to distinguish the peak from the background, rather than a point at which the integrated intensity has dropped smoothly to zero. This points to an important role for RIXS when studying broad low-intensity correlations, where resolving and rejecting the strong inelastic intensity present in soft x-ray scattering, may be vital for resolving weak elastic signals.\cite{Ghiringhelli2012}

\begin{figure}
\includegraphics{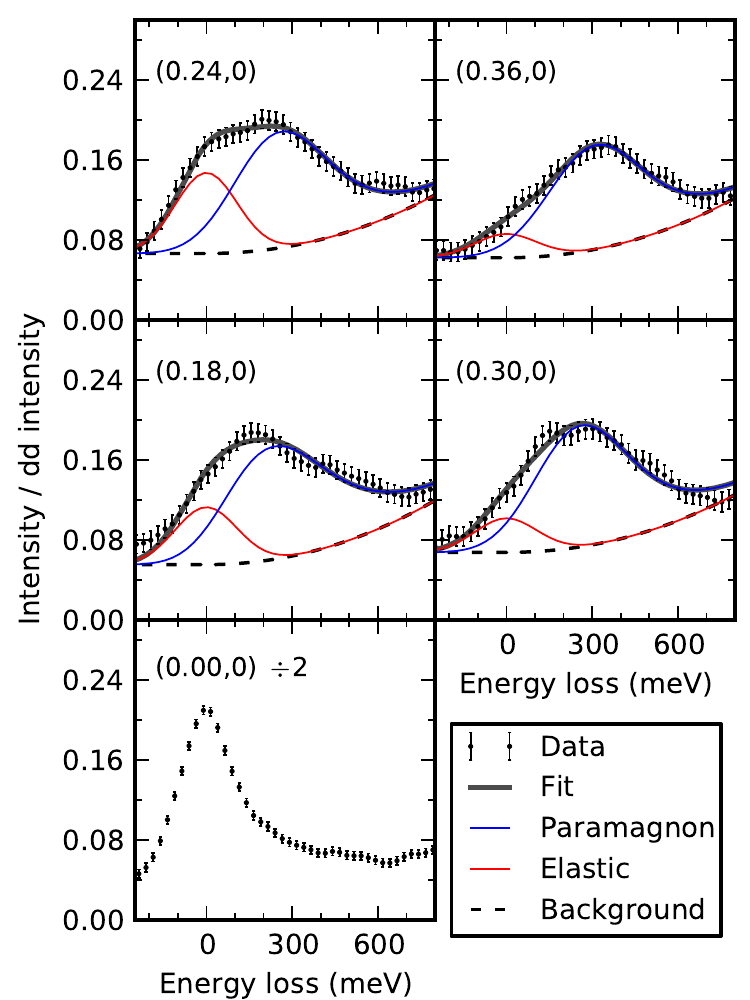} %
\caption{(Color online) RIXS spectra at various \Q{}. Data are plotted with black points with errorbars and the solid gray line represents the fit, which is composed of a paramagnon represented as an anti-symmetrized Lorentizian (blue solid line), elastic intensity (red solid line) and a smooth background (black dotted line). The data were taken at $T=18$~K with $\pi$ polarized x-rays and are presented so that the peak intensity of the $dd$-excitations is one. At $\Q_{\parallel}=0$ the data are dominated by the elastic specular scattering and the spectrum has been divided by a factor of 2.}
\label{Fig3}
\end{figure}

\begin{figure}
\includegraphics{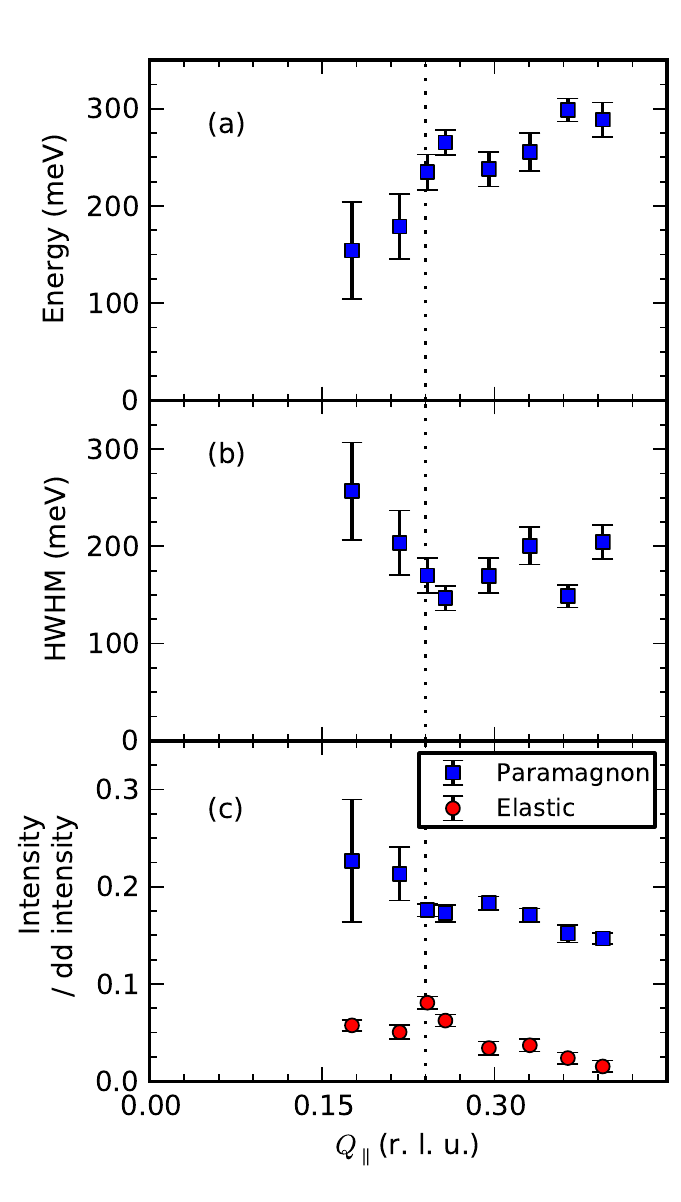} %
\caption{(Color online). The results of fitting the magnetic RIXS dispersion shown in Figs. \ref{Fig2}\&\ref{Fig3} (a) the paramagnon energy, (b) the half-width at half-maximum (HWHM) of the Lorentzian peak and (c) the peak intensity of the elastic line and paramagnon normalized so that the peak intensity of the $dd$-excitations is one at each $Q{_{\parallel}}$. The CO wavevector at $Q{_{\parallel}}=0.24$ is marked by a dotted line.}
\label{Fig4}
\end{figure}

Having characterized the CO in our LBCO 1/8 sample, we now examine the magnetic correlations. Figure~\ref{Fig2} plots a colormap of the excitations at 18~K using $\pi$ incident x-rays in order to enhance the dynamic magnetic scattering. We observe a broad peak, which disperses from low energies at $Q_{\parallel}=0$ up to $\sim 300$~meV near the zone boundary, passing through the CO wavevector at $Q_{\parallel}=0.24$ (marked by an arrow in Fig.~\ref{Fig2}). The stripe wavevector is associated with additional RIXS scattering intensity. In order to analyze the origin of this additional intensity the RIXS spectra we fit a model function to the data in a similar manner to other RIXS studies of doped cuprates. \cite{LeTacon2011, Dean2013b, LeTacon2013}  The elastic scattering is accounted for by a resolution limited Gaussian. The paramagnon is represented by an anti-symmeterized Lorentzian and the background from charge-transfer scattering and the tail of the dd excitations is represented as a smooth line. In order to account for the finite experimental energy resolution, the anti-symmeterized Lorentzian is convolved numerically with a Gaussian of HWHM 130~meV. For $Q_{\parallel} \lesssim 0.16$, the paramagnon energy is too low to unambiguously distinguish between the paramagnon and the elastic and low-energy phonon scattering.  Figure~\ref{Fig3} plots the fits for several values of $Q_{\parallel}$ and Fig.~\ref{Fig4} plots the dispersion of the fitting parameters. At the CO wavevector $Q_{\parallel}=0.24$, the paramagnon fitting shows a slightly higher energy and a small decrease in width. However, within the size of the errorbars, there is no unambiguous change in the energy, width or intensity of the paramagnon. Rather the intensity of the elastic line in Fig.~\ref{Fig4}(c) is seen to increase. So the increase in intensity at $Q_{\parallel}=0.24$ in Fig.~\ref{Fig2} may be due to an increase in the tail of the elastic intensity. Thus, within the errorbars of these measurements the paramagnon appears to be almost independent of the CO, although some uncertainty derives from the decomposition between the paramagnon and elastic intensity. In fact, around the CO wavevector the elastic and paramagnon signal intensities appear to be anticorrelated (Fig.~\ref{Fig4}(c)), and the HWHM and the energy position of the paramagnon show a discontinuity (Fig.~\ref{Fig4}(a\&b)). This could be taken to imply that the magnetic excitations are only minimally affected by the static charge order. However, given that in LBCO the CO forms as a cooperative modulation of both the spins and charge, this seems highly unlikely. We note that the current RIXS resolution (130~meV HWHM) means that only the high-energy magnetic scattering can be resolved in this experiment. It seems likely that changes do occur in the excitation spectrum, but mainly at low energy scales below about 100~meV.

It would be very interesting if this low energy spectrum could be determined in future experiments, such as those being envisaged using the higher resolution RIXS spectrometers the are currently being designed and constructed.  In particular, measuring the magnetic excitations spectrum of LBCO offers the opportunity to identify excitations associated with
the one dimensional nature of the hole-poor magnetic stripe itself as opposed to the two dimensional stripe-ordered magnetic superstructure.  Such features have already been discerned in the related nickelate system La$_{5/3}$Sr$_{1/3}$NiO$_4$, where dispersing magnetic excitations
with a bandwidth of 10~meV were measured.\cite{Boothroyd2003}  Measurements of such excitations in LBCO would permit one to distinguish between bond and site centered stripes\cite{Zaanen1999}, and to determine more generally the distribution of anisotropic holes
within the CuO$_2$ plane.\cite{Seibold2006, Seibold2012} The distribution of dopants in bond and site centered stripe scenarios determines the $Q_{\parallel}$ at which a magnetic response will be seen.  RIXS should also be able to discern possible contributions to the low energy
magnetic response coming from not just hole-poor regions but hole-rich ones as well. While the expectation is that hole-poor regions will provide the strongest magnetic response, RIXS might be able to discern the distinct magnetic signature\cite{Konik2007,Konik2008} of hole-rich regions.

Rather than display an evident signature of the CO, the paramagnon in LBCO 1/8 in our measured \Q{}-range is similar to the magnon observed in other cuprate families such as La$_{2-x}$Sr$_x$CuO$_4$, \cite{Braicovich2009, Braicovich2010, BraicovichPRB2010, Dean2012, Dean2013b} YBa$_2$Cu$_4$O$_8$/YBa$_2$Cu$_3$O$_7$, \cite{LeTacon2011, LeTacon2013} Bi$_2$Sr$_2$CaCu$_2$O$_{8+\delta}$, \cite{Dean2013a} and Tl$_2$Ba$_2$CuO$_{6+\delta}$. \cite{LeTacon2013} Indeed, the HWHM of LBCO 1/8 falls in between the values width reported for La$_{2-x}$Sr$_x$CuO$_4$ with $x=0.11$ and $x=0.16$ consistent with the paramagnon width being a simple function of the hole concentration.

In conclusion, the CO stripe in LBCO 1/8 was observed using Cu $L_3$ edge RIXS and found to have a incident x-ray polarization dependence consistent with charge, rather than spin order. The paramagnon was then measured as it dispersed through this wavevector. Within the resolution of the current measurements, there are no unambiguous changes in the paramagnon due to the CO. We propose that future higher energy resolution experiments are required to observe the coupling between CO and the paramagnon in low energy features below approximately 100~meV such as those being planned at ID32 at the European Synchrotron Radiation Facility, I21 at the Diamond Light Source and at SIX at the National Synchrotron Light Source II.

\begin{acknowledgments}
We thank Jos\'{e} Lorenzana for important discussions and sharing some unpublished results. M.P.M.D., R.M.K.\ and J.P.H.\ are supported by the Center for Emergent Superconductivity, an Energy Frontier Research Center funded by the U.S.\ DOE, Office of Basic Energy Sciences. Work at Brookhaven National Laboratory was supported by the Office of Basic Energy Sciences, Division of Materials Science and Engineering, U.S. Department of Energy under Award No.\ DEAC02-98CH10886. Work by G.G., G.D.\ and M.M.\ was partially supported by the Italian Ministry of Research MIUR (Grant No. PRIN-20094W2LAY). These experiments were performed on the ID08 beamline at the European Synchrotron Radiation Facility (ESRF), Grenoble, France.
\end{acknowledgments}


\begin{thebibliography}{10}

\bibitem{Kivelson2003}
S.~A. Kivelson, I.~P. Bindloss, E. Fradkin, V. Oganesyan, J.~M. Tranquada, A.
  Kapitulnik, and C. Howald, Rev. Mod. Phys. {\bf 75},  1201  (2003).

\bibitem{Vojta2009}
M. Vojta, Advances in Physics {\bf 58},  699  (2009).

\bibitem{Fujita2012stripes}
M. Fujita, Physica C: Superconductivity {\bf 481},  23   (2012).

\bibitem{Tranquada1995}
J.~M. {Tranquada}, B.~J. {Sternlieb}, J.~D. {Axe}, Y. {Nakamura}, and S.
  {Uchida}, \nat {\bf 375},  561  (1995).

\bibitem{Tranquada1997}
J.~M. Tranquada, J.~D. Axe, N. Ichikawa, A.~R. Moodenbaugh, Y. Nakamura, and S.
  Uchida, Phys. Rev. Lett. {\bf 78},  338  (1997).

\bibitem{Ichikawa2000}
N. Ichikawa, S. Uchida, J.~M. Tranquada, T. Niem\"oller, P.~M. Gehring, S.-H.
  Lee, and J.~R. Schneider, Phys. Rev. Lett. {\bf 85},  1738  (2000).

\bibitem{Fujita2004}
M. Fujita, H. Goka, K. Yamada, J.~M. Tranquada, and L.~P. Regnault, Phys. Rev.
  B {\bf 70},  104517  (2004).

\bibitem{Hucker2010}
M. H\"ucker, M. v.~Zimmermann, M. Debessai, J.~S. Schilling, J.~M. Tranquada,
  and G.~D. Gu, Phys. Rev. Lett. {\bf 104},  057004  (2010).

\bibitem{Tranquada2004}
J.~M. {Tranquada}, H. {Woo}, T.~G. {Perring}, H. {Goka}, G.~D. {Gu}, G. {Xu},
  M. {Fujita}, and K. {Yamada}, \nat {\bf 429},  534  (2004).

\bibitem{Xu2007}
G. Xu, J.~M. Tranquada, T.~G. Perring, G.~D. Gu, M. Fujita, and K. Yamada,
  Phys. Rev. B {\bf 76},  014508  (2007).

\bibitem{Boothroyd2003}
A.~T. Boothroyd, P.~G. Freeman, D. Prabhakaran, A. Hiess, M. Enderle, J. Kulda,
  and F. Altorfer, Phys. Rev. Lett. {\bf 91},  257201  (2003).

\bibitem{Moodenbaugh1988}
A.~R. Moodenbaugh, Y. Xu, M. Suenaga, T.~J. Folkerts, and R.~N. Shelton, Phys.
  Rev. B {\bf 38},  4596  (1988).

\bibitem{Emery1997}
V.~J. Emery, S.~A. Kivelson, and O. Zachar, Phys. Rev. B {\bf 56},  6120
  (1997).

\bibitem{Kivelson1998}
S.~A. {Kivelson}, E. {Fradkin}, and V.~J. {Emery}, \nat {\bf 393},  550
  (1998).

\bibitem{Eschrig2006}
M. Eschrig, Advances in Physics {\bf 55},  47  (2006).

\bibitem{Scalapino2012}
D.~J. Scalapino, Rev. Mod. Phys. {\bf 84},  1383  (2012).

\bibitem{Castellani1995}
C. Castellani, C. Di~Castro, and M. Grilli, Phys. Rev. Lett. {\bf 75},  4650
  (1995).

\bibitem{Ghiringhelli2004}
G. Ghiringhelli, N.~B. Brookes, E. Annese, H. Berger, C. Dallera, M. Grioni, L.
  Perfetti, A. Tagliaferri, and L. Braicovich, Phys. Rev. Lett. {\bf 92},
  117406  (2004).

\bibitem{Ghiringhelli2012}
G. Ghiringhelli, M. Le~Tacon, M. Minola, S. Blanco-Canosa, C. Mazzoli, N.~B.
  Brookes, G.~M. De~Luca, A. Frano, D.~G. Hawthorn, F. He, T. Loew, M.~M. Sala,
  D.~C. Peets, M. Salluzzo, E. Schierle, R. Sutarto, G.~A. Sawatzky, E.
  Weschke, B. Keimer, and L. Braicovich, Science {\bf 337},  821  (2012).

\bibitem{Braicovich2009}
L. Braicovich, L.~J.~P. Ament, V. Bisogni, F. Forte, C. Aruta, G. Balestrino,
  N.~B. Brookes, G.~M.~D. Luca, P.~G. Medaglia, F.~M. Granozio, M. Radovic, M.
  Salluzzo, J.~V.~D. Brink, and G. Ghiringhelli, Phys. Rev. Lett. {\bf 167401},
   22  (2009).

\bibitem{Braicovich2010}
L. Braicovich, J. van~den Brink, V. Bisogni, M.~M. Sala, L.~J.~P. Ament, N.~B.
  Brookes, G.~M. {De Luca}, M. Salluzzo, T. Schmitt, V.~N. Strocov, and G.
  Ghiringhelli, Phys. Rev. Lett. {\bf 104},  077002  (2010).

\bibitem{BraicovichPRB2010}
L. Braicovich, M. {Moretti Sala}, L.~J.~P. Ament, V. Bisogni, M. Minola, G.
  Balestrino, D. {Di Castro}, G.~M. {De Luca}, M. Salluzzo, G. Ghiringhelli,
  and J. van~den Brink, Phys. Rev. B {\bf 81},  174533  (2010).

\bibitem{LeTacon2011}
M. Le~Tacon, G. Ghiringhelli, J. Chaloupka, M.~M. Sala, V. Hinkov, M.~W.
  Haverkort, M. Minola, M. Bakr, K.~J. Zhou, S. Blanco-Canosa, C. Monney, Y.~T.
  Song, G.~L. Sun, C.~T. Lin, G.~M. De~Luca, M. Salluzzo, G. Khaliullin, T.
  Schmitt, L. Braicovich, and B. Keimer, Nat. Phys. {\bf 7},  725  (2011).

\bibitem{Dean2012}
M.~P.~M. Dean, R.~S. Springell, C. Monney, K.~J. Zhou, J. Pereiro, I. Bo{\v
  z}ovi{\'c}, B. Dalla~Piazza, H.~M. R{\o}nnow, E. Morenzoni, J. van~den Brink,
  T. Schmitt, and J.~P. Hill, Nat. Mater. {\bf 11},  850  (2012).

\bibitem{Dean2013a}
M.~P.~M. Dean, A.~J.~A. James, R.~S. Springell, X. Liu, C. Monney, K.~J. Zhou,
  R.~M. Konik, J.~S. Wen, Z.~J. Xu, G.~D. Gu, V.~N. Strocov, T. Schmitt, and
  J.~P. Hill, Phys. Rev. Lett. {\bf 110},  147001  (2013).

\bibitem{LeTacon2013}
M. {Le Tacon}, M. {Minola}, D.~C. {Peets}, M. {Moretti Sala}, S.
  {Blanco-Canosa}, V. {Hinkov}, R. {Liang}, D.~A. {Bonn}, W.~N. {Hardy}, C.~T.
  {Lin}, T. {Schmitt}, L. {Braicovich}, G. {Ghiringhelli}, and B. {Keimer},
  ArXiv:1303.3947  (2013).

\bibitem{Dean2013b}
M.~P.~M. {Dean}, G. {Dellea}, R.~S. {Springell}, F. {Yakhou-Harris}, K.
  {Kummer}, N.~B. {Brookes}, X. {Liu}, Y.-J. {Sun}, J. {Strle}, T. {Schmitt},
  L. {Braicovich}, G. {Ghiringhelli}, I. {Bozovic}, and J.~P. {Hill},
  ArXiv:1303.5359  (2013).

\bibitem{Wilkins2011}
S.~B. Wilkins, M.~P.~M. Dean, J. Fink, M. H\"ucker, J. Geck, V. Soltwisch, E.
  Schierle, E. Weschke, G. Gu, S. Uchida, N. Ichikawa, J.~M. Tranquada, and
  J.~P. Hill, Phys. Rev. B {\bf 84},  195101  (2011).

\bibitem{Thampy2013}
v. Thampy \emph{et al.} in preparation.

\bibitem{Ghiringhelli2006_AXES}
M.~E. Dinardo, A. Piazzalunga, L. Braicovich, V. Bisogni, C. Dallera, K.
  Giarda, M. Marcon, A. Tagliaferri, and G. Ghiringhelli, Nucl. Instrum.
  Methods Phys. A {\bf 570},  276  (2007).

\bibitem{Dallera1996}
C. Dallera, E. Puppin, G. Trezzi, N. Incorvaia, A. Fasana, L. Braicovich, N.~B.
  Brookes, and J.~B. Goedkoop, Journal of Synchrotron Radiation {\bf 3},  231
  (1996).

\bibitem{Hucker2013}
M. H\"ucker, M. v.~Zimmermann, Z.~J. Xu, J.~S. Wen, G.~D. Gu, and J.~M.
  Tranquada, Phys. Rev. B {\bf 87},  014501  (2013).

\bibitem{supplementary}
see Supplemental Material at [\emph{insert URL}] for a diagram of the
  experimental geometry.

\bibitem{Ament2011}
L.~J.~P. Ament, M. van Veenendaal, T.~P. Devereaux, J.~P. Hill, and J. van~den
  Brink, Rev. Mod. Phys.  (2011).

\bibitem{Abbamonte2005}
P. {Abbamonte}, A. {Rusydi}, S. {Smadici}, G.~D. {Gu}, G.~A. {Sawatzky}, and
  D.~L. {Feng}, Nature Physics {\bf 1},  155  (2005).

\bibitem{Kim2008}
Y.-J. Kim, G.~D. Gu, T. Gog, and D. Casa, Phys. Rev. B {\bf 77},  064520
  (2008).

\bibitem{Tranquada2008}
J.~M. Tranquada, G.~D. Gu, M. H\"ucker, Q. Jie, H.-J. Kang, R. Klingeler, Q.
  Li, N. Tristan, J.~S. Wen, G.~Y. Xu, Z.~J. Xu, J. Zhou, and M. v.~Zimmermann,
  Phys. Rev. B {\bf 78},  174529  (2008).

\bibitem{Hucker2011}
M. H\"ucker, M. v.~Zimmermann, G.~D. Gu, Z.~J. Xu, J.~S. Wen, G. Xu, H.~J.
  Kang, A. Zheludev, and J.~M. Tranquada, Phys. Rev. B {\bf 83},  104506
  (2011).

\bibitem{Guarise2010}
M. Guarise, B. {Dalla Piazza}, M. {Moretti Sala}, G. Ghiringhelli, L.
  Braicovich, H. Berger, J. Hancock, D. van~der Marel, T. Schmitt, V. Strocov,
  L. Ament, J. van~den Brink, P.-H. Lin, P. Xu, H. R\o{}nnow, and M. Grioni,
  Phys. Rev. Lett. {\bf 105},  157006  (2010).

\bibitem{Ament2009}
L. Ament, G. Ghiringhelli, M. Sala, L. Braicovich, and J. van~den Brink, Phys.
  Rev. Lett. {\bf 103},  117003  (2009).

\bibitem{Haverkort2010}
M.~W. Haverkort, Phys. Rev. Lett. {\bf 105},  167404  (2010).

\bibitem{Igarashi2011}
J.-i. Igarashi and T. Nagao, Phys. Rev. B {\bf 85},  064422  (2012).

\bibitem{Zaanen1999}
J. Tworzyd\l{}o, O.~Y. Osman, C.~N.~A. van Duin, and J. Zaanen, Phys. Rev. B
  {\bf 59},  115  (1999).

\bibitem{Seibold2006}
G. Seibold and J. Lorenzana, Phys. Rev. B {\bf 73},  144515  (2006).

\bibitem{Seibold2012}
G. Seibold, M. Grilli, and J. Lorenzana, Physica C: Superconductivity {\bf
  481},  132   (2012), stripes and Electronic Liquid Crystals in Strongly
  Correlated Materials.

\bibitem{Konik2007}
F.~H.~L. Essler and R.~M. Konik, Phys. Rev. B {\bf 75},  144403  (2007).

\bibitem{Konik2008}
R.~M. Konik, F.~H.~L. Essler, and A.~M. Tsvelik, Phys. Rev. B {\bf 78},  214509
   (2008).

\end{thebibliography}

\end{document}